\documentclass{aastex}

\usepackage{emulateapj5}
\usepackage{psfig}

\def\ltsima{$\; \buildrel < \over \sim \;$}
\def\simlt{\lower.5ex\hbox{\ltsima}}
\def\gtsima{$\; \buildrel > \over \sim \;$}
\def\simgt{\lower.5ex\hbox{\gtsima}}
 
\received{}
\accepted{}
\journalid{}{}
\articleid{}{}
\slugcomment{Accepted for publication in ApJ}
\righthead{DE GRANDI \& MOLENDI}
\lefthead{METALLITICY GRADIENTS IN X-RAY CLUSTERS OF GALAXIES}

\begin{document}
   
\title{Metallicity Gradients in X-ray Clusters of Galaxies}

\author{Sabrina De Grandi}
 \affil{Osservatorio Astronomico di Brera, Via Bianchi 46,
I-23807 Merate (LC), Italy}
\email{degrandi@merate.mi.astro.it}

\and

\author{Silvano Molendi}
 \affil{Istituto di Fisica Cosmica, CNR, via Bassini 15,
I-20133 Milano, Italy}
\email{silvano@ifctr.mi.cnr.it}

\begin{abstract}
We present the projected metallicity profiles for a sample of 17 rich 
galaxy clusters observed by BeppoSAX.
We find that the 8 non-cooling flow clusters have flat metallicity
profiles. On the contrary, a strong enhancement in the abundance is
found in the central regions of the cooling flow clusters.
All the non-cooling flow clusters present evidence of recent merger
activity suggesting that the merger events redistributes efficiently
the metal content of the intracluster medium.
For the cooling flow clusters with better statistics and available 
optical data (A85, A496, A2029 and Perseus) we have tested whether 
the observed abundance excess is due to metals ejected from the 
galaxies located in the cluster core.
We find that at a resolution $\simgt 100$ kpc the observed projected
abundance excess profiles are consistent with originating from a deprojected
metal excess distribution tracing the optical light distribution.
In the one case (i.e. Perseus) with higher resolution ($\sim 50$ kpc),
we find that the observed metal abundance excess distribution is broader than
the predicted one.
Such a difference can be reconciled if we assume that the metals have
drifted away from their parent ejecting galaxies by a few tens of kpc,
or, alternatively, if we assume that the cluster light profile has
become significantly more centrally peaked because of the formation 
process of the central dominant cluster galaxy since the last major 
merger occurred.
\end{abstract}


\section {Introduction}

The X-ray spectra of clusters of galaxies is rich in emission lines
due to K and L-shell transitions from highly ionized heavy elements,
the most prominent is generated by iron at about 6.9 keV.  From the
measure of the equivalent width of the lines it is possible to derive
the abundance of the emitting elements in the gas.  For rich clusters
between redshift 0.3 and the present day the observed abundance of
iron is about 1/3 the solar value (e.g., Mushotzky \& Loewenstein 1997,
Fukazawa et al. 1998, Allen \& Fabian 1998 (AF98)), suggesting that a
significant fraction of the intracluster medium (ICM) has been
processed into stars already at intermediate redshift.

Arnaud et al. (1992) showed that the iron mass in the ICM is highly
correlated with the optical light from ellipticals and lenticulars
suggesting that early-type galaxies are the major source of the
enrichment of the ICM.
The metal enrichment mechanisms for the ICM remain controversial.
Mushotzky et al. (1996) and Mushotzky \& Loewenstein (1997) showed
that the relative abundance of elements is consistent with an origin
of all the metals in Type II supernovae (SN), supporting the
proto-galactic winds scenario for the metal enrichment of the ICM.
However, other works on ASCA data (Ishimaru \& Arimoto 1997, Fukazawa
et al. 1998, Finoguenov, David \& Ponman 2000, Dupke \& White 2000),
still indicating a predominance of the Type II SNe enrichment at large
radii in clusters, do not exclude that as much as $50\%$ of the iron
in clusters could comes from Type Ia SNe ejecta in the inner parts of
the clusters.  Very recent observations with XMM EPIC (B\"ohringer et
al. 2000; Tamura et al. 2000) provide strong evidence in favor of a
substantial contribution of SN Ia in the enrichment of the ICM.

X-ray observations show that in the central region of $70-90$ per cent
of the clusters the cooling time of the ICM is less than a Hubble time
(e.g., Peres et al. 1998 (P98)) and that, under gravity and thermal
pressure, the gas cools and flows inwards to maintain pressure
equilibrium. This process is known as a cooling flow (see Fabian 1994
for a review).
An interesting result found by AF98 shows that the metallicity depends
on whether or not a cluster has a cooling flow. In particular, cooling
flow clusters have metallicity $\sim 1.8$ times higher than that of
the non-cooling flow systems.  AF98 suggest that this is caused by the
presence of metallicity gradients in the cooling flow clusters.

Spatially resolved abundance measurements in galaxy clusters are of
great importance because they can be used to measure the precise
amount of metals in the ICM
and to constrain the origin of metals both spatially and
in terms of the contribution of different type of SNe as a function of
the position in the cluster.
Abundance gradients have been measured for a few clusters.  Early
works on individual objects include that on the Centaurus cluster
(Fukazawa et al 1994), Hydra A (Ikebe et al. 1997), AWM7 (Ezawa et
al. 1997), Perseus cluster (Arnaud et al. 1994, Molendi et al. 1998),
and Virgo (Matsumoto et al. 1996, Guainazzi \& Molendi 1999).  Later works on
cluster samples from ASCA data are those of White (2000) on a sample
of 106 clusters, Dupke \& White (2000) on 3 clusters, and Finoguenov
et al. (2000) on 11 clusters.
  
A combination of good angular and energy resolution is required to
resolve the metallicity distribution.  The spectral range of ASCA
reaches energies up to 10 keV, which are typical for hot clusters, but
its point-spread function (PSF) has a large ($\sim 2^\prime$)
half-power radius (HPR) that depends strongly upon the energy
(Serlemitsos et al. 1995).
A better combination of the properties required is provided by the
Medium Energy Concentrator Spectrometer (MECS, Boella et al. 1997a) on
board the BeppoSAX satellite (Boella et al. 1997b) working in the 1-10
keV with a PSF of $\sim 1^\prime$ (HPR), varying only weakly with the
energy (D'Acri, De Grandi \& Molendi 1998), and a spectral resolution
of $\sim 8\%$ at 6 keV.

In this paper we present spatially resolved metal abundance
measurements for a sample of 17 rich and nearby ($z\lesssim 0.1$)
X-ray clusters observed with BeppoSAX.  These are all the clusters
with on-axis pointings and exposure times larger than 30 ks which were
available at the BeppoSAX SDC archive at the end of August 2000.  
The observation log for the cluster sample is given in Table 1.

We have already published the metallicity measurements for some of
these clusters (Perseus, Molendi et al. 1998; A2319, Molendi et
al. 1999; A2029, Molendi \& De Grandi 1999; PKS0745$-$191, De Grandi
\& Molendi 1999a; A3266, De Grandi \& Molendi 1999b; A2256, Molendi, 
De Grandi \& Fusco-Femiano 2000; A3562, Ettori et al. 2000).
Here, for the first time, we explore systematically the relationship
between metallicity gradients in the ICM in clusters of galaxies and
the presence of a cooling flow.

Very recently, after this paper was submitted we have learned that
Irwin \& Bregman (2000) obtained metallicity gradients for a sample of
12 galaxy clusters observed by BeppoSAX. Their work differs from ours
in three major ways: their sample is smaller than ours; they study
profiles out to 9 arcmin only from the cluster core while we extend
our analysis further out; and, finally, they do not investigate the
difference between cooling flow and non-cooling flow clusters in such
a systematic fashion as we do.

Unless otherwise stated, all uncertainties quoted in this paper are
68$\%$ confidence levels for 1 interesting parameter ($\Delta \chi^2 =
1$). 
The abundances estimated are all relative to the cosmic values given
in Anders \& Grevesse (1989).  Throughout this paper, we assume $H_0 =
50$ km s$^{-1}$ Mpc$^{-1}$, $\Omega=1$ and $\Lambda=0$.

\section {Observation and Data Analysis}

Standard reduction procedures and screening criteria have been applied
using SAXDAX package under FTOOLS environment,
to produce equalized and linearized MECS event files.  Moreover all events 
occurring at times when the instantaneous pointing direction differs
by more than 10$^{\prime\prime}$ from the mean pointing
direction have been rejected.

Albeit small, the PSF-induced spectral distortions (D'Acri et al. 1998)
have been taken into account using appropriated effective area files
produced with the EFFAREA program, as described in
Molendi et al. (1999).
All MECS spectra have been background subtracted using spectra
extracted from blank sky event files in the same region of the
detector as the source. 
All spectral fits have been performed using XSPEC Ver. 10.00.

\medskip
\psfig{file=f1.eps,width=3.in,angle=-90}

\figcaption
{MECS spectra extracted from five concentric annular regions in the
cooling flow cluster A85. Minimum and maximum bounding radii for the
five regions are reported in each panel. The Fe-K$_\alpha$ emission
line at 6.9 keV declines rapidly with the radius.}
\medskip

Each cluster has been divided into concentric annuli centered on the
X-ray emission peak; out to 8$^\prime$ we accumulate spectra from 4
annular regions each 2$^\prime$ wide, beyond this radius we accumulate
spectra from annuli 4$^\prime$ wide.
For each cluster the radial profile stops at the last annulus were the
intensity of the source at about 7 keV is roughly equal to that of the
background.
The range adopted for spectral fitting is 2-10 keV except for the 
8$^\prime$-12$^\prime$ annulus, where the
correction for the strongback is most important. Here we restrict ourselves
to the range 3.5-10 keV to avoid the low energy part of the spectrum
where the correction is less reliable.

The dominant contribution to the MECS background at energies larger
than $\sim$ 5 keV is from events induced by the interaction of high
energy particles with the structure surrounding the instrument. Using
data acquired during occultations of the satellite from the dark earth,
Vecchi et al. (1999) have monitored the non X-ray background finding
that variations are typically contained within $\sim 5\%$ from the
mean.  In the present work, we have decided to account for these
variations by excluding from our analysis spectra from the outermost
regions not satisfying the conditions on the intensity of the source
with respect to the background indicated in the previous paragraph.
We believe that this is preferable to the alternative choice of using
this data and including a systematic component to the error budget to
account for possible variations in the background, the reason being
that if such a component exists and is truly systematic it will show
up again when we average our metallicity profiles (see Fig. 4).
For the spectra which satisfy the conditions on the intensity of the
source with respect to the background indicated in the previous
paragraph, fluctuations of the background of up to $\sim 5\%$ do not
introduce a quantifiable increase in the error of the abundance
measurement.

We have divided our clusters into a subsample of non-cooling-flow
(non-CF) and cooling-flow (CF) objects on the basis of the ROSAT
analysis presented in P98.  For the non-CF systems we fit each
spectrum with an isothermal plasma in collisional equilibrium at the
redshift of the cluster ({\it mekal} model in XSPEC), absorbed by the
nominal Galactic column density (Dickey \& Lockman 1990; {\it wabs}
model). The free parameters are the temperature ($kT$) and metallicity
($Z$) of the gas and the normalization.
 
Given the adopted spectral range (i.e. E $>2$ keV in all cases) the
most important emission line complex we observe is the Fe K$_\alpha$
line blend at 7 keV. This complex is the only one to clearly stand out
from visual inspection of the observed spectra (see Fig. 1).  The
reasons for the dominance of this complex over others are several:
assuming solar ratios, Fe is one of the ten most abundant elements
(e.g., Anders \& Grevesse 1989); the complex is well isolated from
other emission features; and, finally, it is located at an energy
where the MECS have a good spectral resolution and the effective areas
are still relatively large. We note that although the spectral
resolution of the MECS is insufficient to resolve the Fe He-like
K$_\alpha$ line at 6.7 keV from the Fe H-like K$_\alpha$ line at 6.9
keV, it is sufficient to clearly separate the line blend from the
continuum emission (see Fig. 1), and therefore derive a robust
measurement of the equivalent width of the line, which, of course, is
what is required to determine the Fe abundance.  From the above
discussion it follows that our metallicity measurements are dominated
by Fe and that they cannot be used to discriminate between enrichment
entirely dominated by type II Supernovae (e.g., Mushotzky et al. 1996)
or enrichment from a mix of Type II and Type Ia Supernovae (e.g.,
Dupke
\& White 2000).

For the innermost region of CF clusters we have performed spectral
fits with two distinct emission models: the first is the single
temperature model applied to all other regions (i.e. {\it mekal}
model); the second includes the single temperature component plus an
additional component modeling the emission from the cooling flow. This
second component ({\it mkcflow} model) accounts for a multiphase
atmosphere, which cools at constant pressure from the ambient cluster
temperature.  All parameters of the cooling flow component, except for
the normalization parameterized in terms of a mass deposition rate,
are constrained: the minimum temperature is fixed at 0.1 keV, the
maximum temperature and the metallicity have been set to be equal to
the temperature and metallicity of the isothermal ambient ({\it
mekal}) component.  We find no evidence for a significant variation in
the metallicity if the mass deposition rate is fixed to values derived
from the literature either from deprojection or spectral analysis
(P98, Sarazin et al. 1998, Allen 2000).

In table 2 we report the abundance measurements derived for the
$0^\prime-2^\prime$ annular region in both cases.  We find that no
systematic trend is present when comparing the abundances measured
with a single temperature model with those obtained adding a cooling
flow spectral component, and, that the two metallicities are always in
agreement within $\sim 1\sigma$. 

In general the cooling radius estimated for our CF clusters by P98, is
contained within the first 0$^\prime$-2$^\prime$ bin. An exception is
the case of the Perseus cluster with a cooling radius of $\sim
6^\prime$ (P98).  In this specific case we have evaluated the
influence of the cooling flow on the abundances measurement in the bin
0$^\prime$-2$^\prime$, 2$^\prime$-4$^\prime$, 4$^\prime$-6$^\prime$
and 6$^\prime$-8$^\prime$, finding that the metallicities derived
using a {\it mekal} code with or without an additional cooling flow
spectral component ({\it mkcflow}) are again always in agreement
within $1 \sigma$.

In the following we use the abundances estimated from the single
temperature model ({\it mekal} code) in any annular region.  The
metallicity measurements for the 17 clusters are reported in Table 2.

\section {Spatially resolved metallicity measurements} 

For each cluster we compute the mean metal abundance by fitting the
radial profile with a constant. The results are reported in Table 2.
The mean metallicity for the whole sample of 17 clusters is $0.29\pm
0.01$ , in good agreement with previous works (Edge \& Stewart 1991,
Fukazawa et al. 1998, AF98).  The mean metallicity for the CF
subsample (9 clusters) is $0.34\pm 0.01$ , while for the non-CF
subsample (8 objects) it is $0.25\pm 0.01$.

Figure 2 and 3 show the radial abundance profile for the CF and non-CF
clusters, respectively. The abundances are plotted against the radius
in units of $r_{180}$, this is the radius within which the mean
enclosed density equals 180 times the critical density of the universe
and approximates the virial radius of the cluster in an $\Omega=1$
cosmology.  We compute $r_{180}$ as in Markevitch et al. (1998) which
follows the prediction given by the simulations of Evrard, Metzler \&
Navarro (1996): $r_{180}$ = 3.95 Mpc (T$_{X}/10$ keV)$^{1/2}$.  Here
$T_{X}$ is the mean emission-weighted temperature derived from the
temperature profile up to the maximum radius available of each
cluster, in units of keV. 
The values of $r_{180}$ are reported in Table 2.

\centerline{\psfig{file=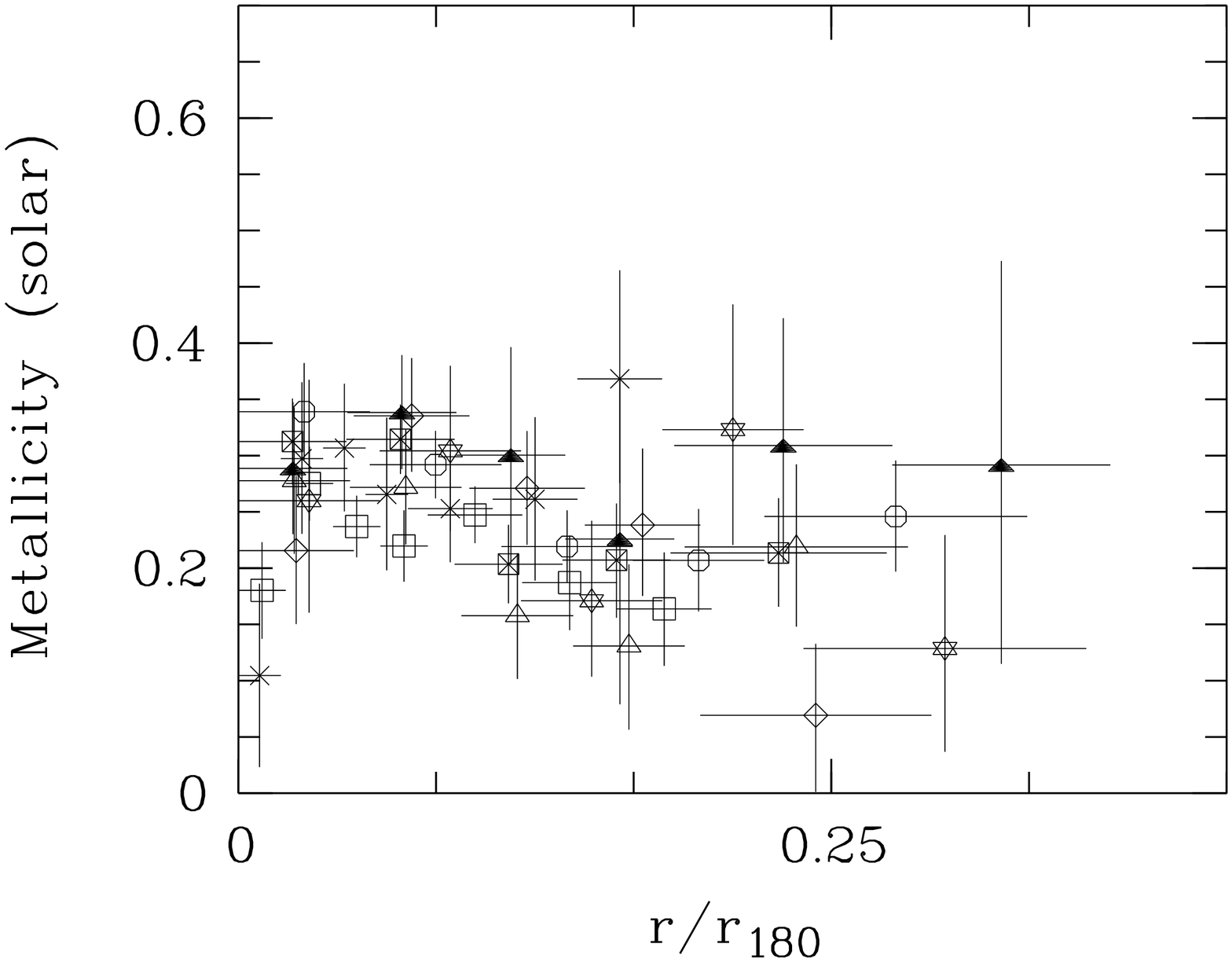,width=3.9in,angle=-90}}

\figcaption
{Metallicity profiles (projected) for the non-CF clusters, plotted
against radii in units of $r_{180}$.  Clusters are related to symbols
as follows: A119 (lozenges), A754 (crossed squares), A2256 (circles),
A2319 (filled triangles), A3266 (triangles), A3376 (stars), A3627
(crosses) and Coma (squares).}
\medskip

The projected abundance profiles of all the non-CF systems, plotted in
Figure 2, are consistent with being constant with the radius (see
Table 2).  All our non-CF clusters present evidences of merger
activity and/or a non-relaxed morphology in X-rays ( e.g., Fabricant
et al. 1993, Mohr et al. 1995, Buote \& Tsai 1996, Buote \& Canizares
1996, Godlowski et al. 1998, Markevitch et al. 1998, Trevese et
al. 2000, Flores et al. 2000).  It seems plausible that in these
clusters mergers have remixed the ICM thereby washing out any
pre-existing metallicity gradient.

\centerline{\psfig{file=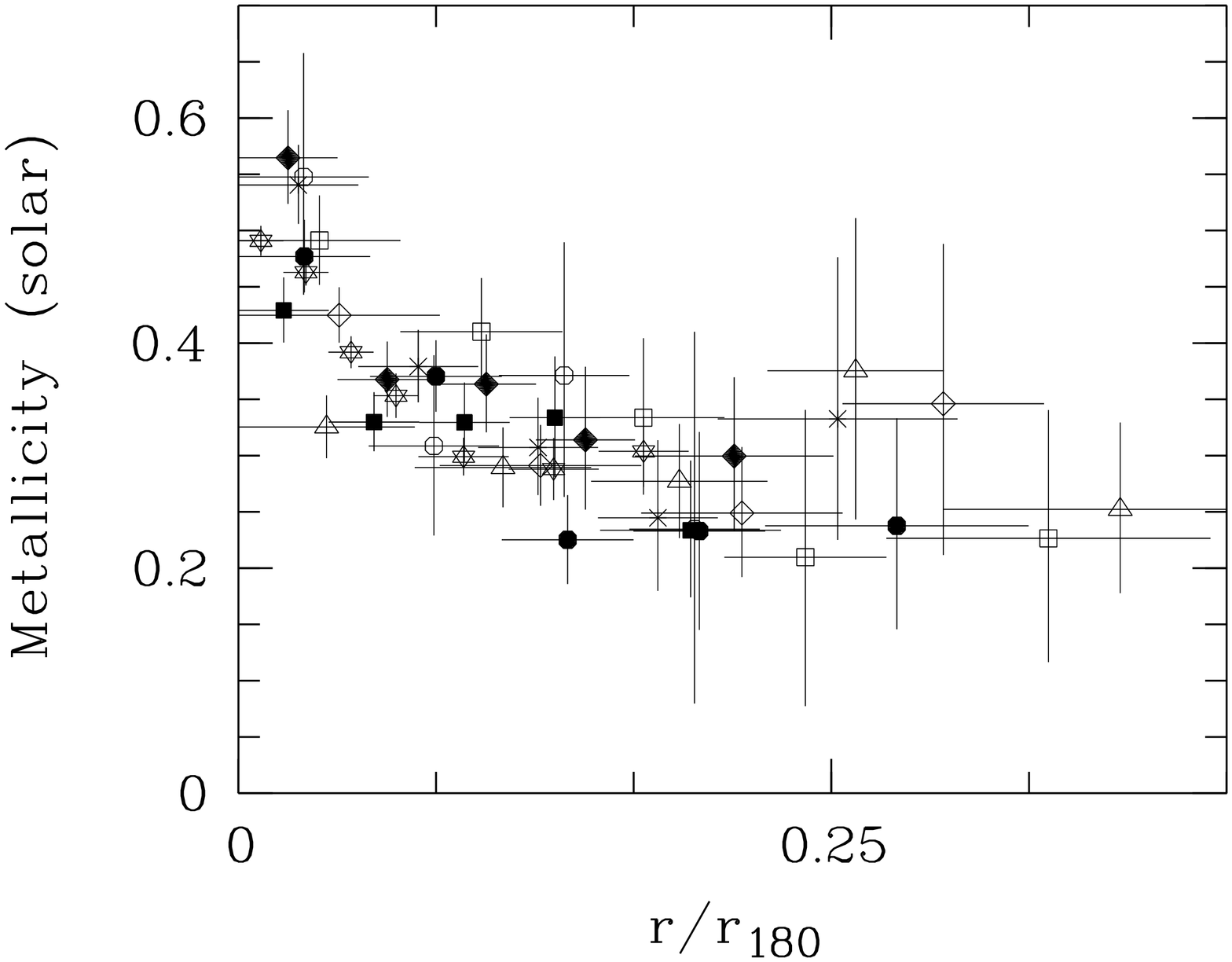,width=3.9in,angle=-90}}

\figcaption
{Metallicity profiles (projected) for the CF clusters, plotted against
radii in units of $r_{180}$. Clusters are related to symbols as
follows: A85 (filled circles), A496 (filled lozenges), Perseus
(stars), A2029 (open squares), A2142 (open triangles), A2199 (filled
squares), A3562 (open circles), 2A0335 (crosses) and PKS0745 (open
lozenges).}
\medskip

The picture for the metallicity profiles of the CF clusters, shown in
Figure 3, is completely different.  A clear evidence of an abundance
gradient declining with the radius is present in most of these
clusters (see Table 2).
Using the $\chi^2$ statistics, a constant provides an acceptable fit
only to the abundance profile for A2142 (probability of 0.78) and
A3562 (probability of 0.29).
A2142 is a well known cooling flow cluster with a cooling radius of
about 150 kpc and a mass deposition rate $> 300$ M$_\odot$ yr$^{-1}$
(P98, Allen 2000). However, several observations at optical (Oegerle,
Hill \& Fitchett 1995) and X-ray (Buote \& Tsai 1996, Henry \& Briel
1996, Markevitch et al. 2000) wavelengths indicate that this cluster
is not in a dynamically relaxed state but experienced a merger event
which probably failed to penetrate the core of A2142.
On the other hand, for A3562 the poor statistical quality of the
data is such that its abundance profile is consistent both with being
constant and with having a declining profile similar to the one
observed in the other CF systems.

Our individual profiles are in broad agreement with those measured by
ASCA (e.g., White 2000).  The most striking difference between the
BeppoSAX and the ASCA results is the substantially better resolution
of the former, allowing us for the first time to place tight
constrains on the shape of most profiles.  The most important reasons
for this difference are: the better spatial resolution of the MECS and
the longer exposure times of the BeppoSAX clusters (see Table 1).

\section {Discussion}

In Figure 4 we compare the mean error-weighted abundance profile for
CF and non-CF clusters.  The metal abundances of the CF clusters are
larger than 0.4 of the solar value in the central regions and decrease
rapidly to values similar to those of the non-CF clusters at radii
$\simgt 0.25 \rm {r_{180}}$.  
The profile for non-CF clusters is much flatter, a fit with a constant
to all non-CF abundance measurements is statistically acceptable
($\chi^2 = 58.8$ for 44 degrees of freedom).
However, a small gradient is present in the data, indeed by performing
a fit with a constant plus a linear component we find a statistically
significant improvement (more than $99.5\%$ level) according to the
F-test.

The comparison of the abundance profile for CF and non-CF clusters
supports the scenario where major merger events disrupt the central
regions of clusters thereby re-mixing the gas within these regions and
therefore destroying pre-existing abundance gradients.  The modest
gradient observed in non-CF clusters is quite likely the relic of a
much stronger gradient which has not been completely wiped out by
merger events.

\centerline{\psfig{file=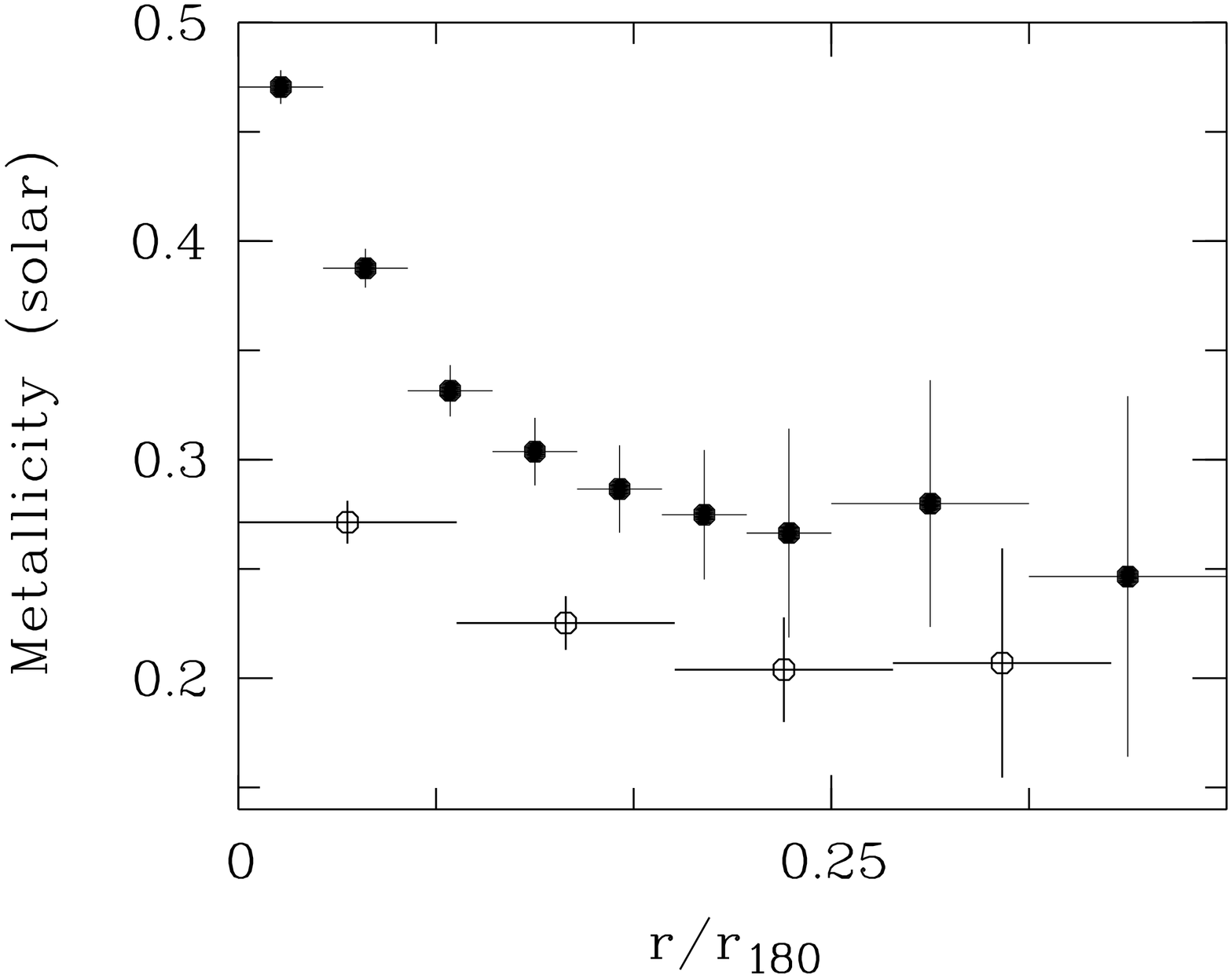,width=3.9in,angle=-90}}

\figcaption
{Mean metallicity profiles for the CF (filled circles) and
non-CF clusters (open circles), plotted against radii in units of
$r_{180}$.}
\medskip

An example of this mechanism can be found in A2256 (see Molendi et
al. 2000).  In this cluster a relatively recent merger event has
disrupted a cooling flow that had already developed in the core of the
infalling subgroups (Fabian \& Daines 1991). Since the merger in A2256
is still in a relatively early stage, the remixing of the gas of the
two subclusters is not complete, thereby generating an abundance
gradient in the direction opposite to the merging direction.

If each merger events redistributes efficiently the metals within the
ICM then the metal excess we see in the core of CF clusters should be
directly related to the enrichment processes which have occurred in
the cluster core since the last major merger. Thus, just as the global
metallicity of clusters is an indicator of the global star formation
history within the whole cluster, the abundance excess we see in the
core of CF clusters is an indicator of the star formation history in
the core of the cluster since the last major merger.

In the light of the above statement, we have tried to test whether the
metal abundance excess we see is due to metals expelled from early
type galaxies located in the core of the cluster. More specifically we
have computed the metal abundance excess profile expected when the metal
excess distribution traces the light distribution of early type galaxies.  It
can be readily shown that the projected abundance excess $Z_{proj}(b_{min},
b_{max})$ measured within a bin with bounding radii $b_{min}$,
$b_{max}$ is related to the deprojected metal abundance excess distribution
$Z(r)$ by the following equation:
\vskip -0.2cm

$$ Z_{proj}(b_{min},b_{max}) = { \int_{b_{min}}^{b_{max}} b~db
\int_{b^2}^{\infty} {n^2(r) Z(r) \over \sqrt{r^2 -b^2} } dr^2 
\over \int_{b_{min}}^{b_{max}} b~db  
\int_{b^2}^{\infty} {n^2(r)\over \sqrt{r^2 -b^2} } dr^2 }. \eqno(1)$$
\smallskip

The abundance excess is proportional to the ratio of the metal density excess, 
in our case iron, $Fe(r)$, to the gas density $n(r)$, i.e. $ Z(r) \propto
Fe(r)/n(r)$. If we assume that the iron excess distribution follows the
distribution of light from early type galaxies, $l(r)$, we have $ Z(r)
\propto l(r)/n(r)$.  

We have computed $Z_{proj}$ for the 4 cooling flow objects where the
metal abundance profile is best measured and optical data is
available, namely: A85, A496, A2029 and Perseus (see Figure 5).  X-ray
and optical light density profiles $n(r)$ and $l(r)$ have been derived
by deprojecting X-ray and optical surface brightness profiles taken
from the literature.  Note that we employ the total optical light
profile including both the light profile of the cD and that of other
early type galaxies in the core of the cluster.

All X-ray profiles are from Mohr et al. (1999), while the optical data
comes from a number of works (A85: Porter et al. 1991, Slezak et
al. 1998; A496: Schombert 1986, Slezak et al. 1999; A2029: Uson et
al. 1991; Perseus: Schombert 1986, Brunzendorf \& Meusinger 1999).
Since our innermost radial bins oversample the core of the MECS PSF by
a factor of 2 only, equation (1) needs to be corrected for the
smoothing effects of the PSF. This is achieved by substituting the
innermost integrals at the numerator and denominator with their
convolutions with the MECS PSF.

In all cases the predicted abundance excess profiles show a central
peak which is due to the cD galaxy. Indeed if we subtract the cD light profile
from the total light profile the derived metal abundance excess profile
becomes flat or even increasing with radius.

For A496 and A2029 the predicted projected metal abundance excess can
be reconciled with the observed one, this is not the case for A85 and
Perseus. For the latter two clusters the predicted projected metal
abundance excess appears to be more centrally concentrated than the
observed one. In the case of A85 the difference is not highly
significant, while for Perseus the two profiles are clearly very
different. We note that Perseus is the nearest cluster in our
sample, if we were to move it to the redshift of the second nearest
object, i.e. A496, the observed and predicted abundance excess profiles would
be in broad agreement.  Thus we conclude that at the resolution of a
few 100 kpc the observed projected abundance excess profiles are consistent
with originating from a deprojected metal excess distribution following the
optical light distribution.  For the one case (i.e. Perseus) where our
resolution is better than 100 kpc we find that the observed abundance
excess distribution is broader than the predicted one.

One possible way of reconciling the observed and predicted abundance
excess profile of Perseus is to assume that metals ejected from elliptical
galaxies are blown far away from them. As pointed out by Ezawa et
al. (1997), the difficulty with this interpretation is that heavy ions
move slowly, the typical distance covered within a Hubble time should
be of the order of a few tens of kpc only.  The predicted metal
abundance excess profile for Perseus could be reconciled with the observed
one if the metals were to drift by about 50 kpc, this is a rather
large but perhaps not impossible drift.
Alternatively, if we assume that the metals have not moved far from
where they have been injected in the ICM we can still explain the
difference between observed and predicted profiles by noting that the
optical luminosity profile we observe is the current luminosity
profile, while the metal excess profile should be related to the optical
luminosity profile averaged over the last few Gyrs since the cluster
suffered the last major merger.  Galaxies that are located at the
center of clusters are thought to be formed by stripping of galaxies
which fall within their gravitational attraction, therefore it is
quite plausible that during such a process the light profile of a
cluster has become more and more centrally concentrated.
This process will inevitably lead to a current light profile which 
is more peaked than the light profile averaged over the time 
period since the last major merger and, as a consequence of this,
the metal abundance excess profile predicted from the current optical 
light profile will be more centrally concentrated than the observed one.

\bigskip
\centerline{\psfig{file=f5.eps,width=3.in,angle=-90}}

\figcaption
{Predicted (dashed lines) versus measured (solid circles) metallicity
excess profiles as a function of radius for the cooling flow clusters
A496, A2029, A85 and Perseus.}
\medskip

\acknowledgments
We thank R. Fusco-Femiano for allowing us to use proprietary data (one
observation of A754 and A119) prior to publication and the anonymous
referee for his useful comments and suggestions which helped us in
improving the paper.  We acknowledge support from the BeppoSAX
Science Data Center.  Part of the software used in this work is based
on the NASA/HEASARC FTOOLS and XANADU packages.









\clearpage


\begin{deluxetable}{lrrllr}
\tabletypesize{\scriptsize}
\tablewidth{0pt}
\tablecolumns{6}
\tablecaption{Observation log for the BeppoSAX cluster sample\tablenotemark{a}}
\tablehead{
\colhead{Target Name} & \colhead{RA(2000)} & \colhead{DEC(2000)} & \colhead{Obs.Date} &
\colhead{Obs.Code} & Duration\nl
\colhead{} & \colhead{(degree)} & \colhead{(degree)} & \colhead{yy-mm-dd} &
\colhead{} & \colhead{(ks)} \nl
}
\startdata 
A85            &  10.3750 &  -9.3833 & 1998-07-18 & 60632001 &  93 \nl 
A119           &  14.0667 &  -1.2494 & 2000-07-05 & 61091002 & 128 \nl 
A426 (Perseus) &  49.9550 &  41.5075 & 1996-09-19 & 60009001 &  80 \nl 
A496           &  68.4071 & -13.2619 & 1998-03-05 & 60477001 &  92 \nl 
A754           & 137.3421 &  -9.6878 & 2000-05-06 & 60936001 &  62 \nl 
               & 137.3375 &  -9.6900 & 2000-05-17 & 61091001 & 123 \nl
A1656 (Coma)   & 194.8950 &  27.9450 & 1997-12-28 & 60126002 &  68 \nl 
               & 194.8950 &  27.9450 & 1998-01-19 & 601260021&  24 \nl 
A2029          & 227.7313 &   5.7439 & 1998-02-04 & 60226001 &  42 \nl
A2142          & 239.5833 &  27.2333 & 1997-08-26 & 60169002 & 102 \nl
A2199          & 247.1592 &  39.5514 & 1997-04-21 & 60169001 & 101 \nl 
A2256          & 255.9929 &  78.6419 & 1998-02-11 & 60465001 &  81 \nl 
               & 255.9929 &  78.6419 & 1999-02-25 & 60126003 &  51 \nl 
A2319          & 290.3025 &  43.9494 & 1997-05-16 & 60226002 &  40 \nl 
A3266          &  67.8379 & -61.4444 & 1998-03-24 & 60539002 &  76 \nl 
A3376          &  90.4058 & -39.9903 & 1999-10-17 & 60936002 & 110 \nl 
A3562          & 203.4100 & -31.6700 & 1999-01-31 & 60638001 &  46 \nl 
A3627          & 243.5917 & -60.8722 & 1997-03-01 & 60180001 &  34 \nl 
2A 0335$+$096  &  54.6458 &   9.9650 & 1998-09-11 & 60675001 & 105 \nl 
PKS 0745$-$191 & 116.8792 & -19.2958 & 1998-10-23 & 60539001 &  92 \nl  
\enddata

\tablenotetext{a}{Multiple observations of the same cluster have been merged.}

\end{deluxetable}

\clearpage


\begin{deluxetable}{lcccccccccc}
\tabletypesize{\scriptsize}
\tablewidth{0pt}
\tablecolumns{11}
\rotate
\tablecaption{Summary of the BeppoSAX MECS metallicity radial profiles, $r_{180}$ and mean metallicity best-fit results.}
\tablehead{
\colhead{Name} & \colhead{Z(0$^{\prime}$-2$^{\prime}$)} & \colhead{Z(2$^{\prime}$-4$^{\prime}$)} & \colhead{Z(4$^{\prime}$-6$^{\prime}$)} & 
\colhead{Z(6$^{\prime}$-8$^{\prime}$)} & \colhead{Z(8$^{\prime}$-12$^{\prime}$)} & \colhead{Z(12$^{\prime}$-16$^{\prime}$)} & 
\colhead{Z(16$^{\prime}$-20$^{\prime}$)} & \colhead{r$_{180}$} & \colhead{$<Z>$} & \colhead{$\chi^2$/dof} \nl
\colhead{}  & \colhead{(solar)} & \colhead{(solar)} & \colhead{(solar)} & \colhead{(solar)} & \colhead{(solar)} & \colhead{(solar)} & \colhead{(solar)} 
  & \colhead{(Mpc)} & \colhead{(solar)} }
\startdata 
CF & & & & & & & & & & \nl 
A85\tablenotemark{a}  & 0.48$\pm$0.03 (0.49$\pm$0.04) & 0.37$\pm$0.03 & 0.23$\pm$0.04 & 0.23$\pm$0.09 & 0.24$^{+0.08}_{-0.07}$ & --- & --- & 3.22 & 0.36$\pm$0.02 & 28.4/4 \nl 

A426 (Perseus) &  0.49$\pm$0.01 (0.48$\pm$0.01) & 0.46$\pm$0.01 & 0.39$\pm$0.01 & 0.35$\pm$0.02 & 0.30$\pm$0.02 & 0.29$\pm$0.03 & 0.30$\pm$0.03 & 3.19 & 0.41$\pm$0.01 & 141.1/6 \nl 

A496           &  0.56$\pm$0.04 (0.53$\pm$0.04) & 0.37$\pm$0.03 & 0.36$\pm$0.04 & 0.31$\pm$0.06 & 0.30$\pm$0.07 & --- & --- & 2.59 & 0.40$\pm$0.02 & 20.4/4 \nl 

A2029          &  0.49$\pm$0.04 (0.50$\pm$0.04) & 0.41$\pm$0.05 & 0.33$\pm$0.07 & 0.21$\pm$0.13 & 0.23$\pm$0.11 & --- & --- & 3.44 & 0.42$\pm$0.03 & 10.2/4 \nl 

A2142\tablenotemark{b} &  0.33$\pm$0.03 (0.33$\pm$0.03) & 0.29$\pm$0.04 & 0.28$\pm$0.05 & 0.38$^{+0.14}_{-0.13}$ & 0.25$^{+0.08}_{-0.07}$ & --- & --- & 3.63 & 0.30$\pm$0.02 & 1.8/4 \nl 

A2199 &   0.43$\pm$0.03 (0.42$\pm$0.03) & 0.33$\pm$0.03 & 0.33$^{+0.04}_{-0.03}$ & 0.33$\pm$0.05 & 0.23$\pm$0.06 & --- & --- & 2.65 & 0.35$\pm$0.02 & 11.8/4 \nl 

A3562\tablenotemark{c}   &  0.55$^{+0.11}_{-0.10}$ (0.55$^{+0.15}_{-0.11}$) & 0.31$\pm$0.08 & 0.37$^{+0.12}_{-0.11}$ & 0.23$^{+0.18}_{-0.15}$ & --- & --- & --- & 2.82 & 0.37$\pm$0.05 & 3.8/3 \nl 

2A 0335$+$096 &  0.54$^{+0.04}_{-0.03}$ (0.47$\pm$0.04) & 0.38$\pm$0.03 & 0.31$\pm$0.04 & 0.24$^{+0.06}_{-0.07}$ & 0.33$^{+0.14}_{-0.11}$ & --- & --- & 2.27 & 0.40$\pm$0.02 & 25.6/4 \nl 

PKS 0745$-$191 &  0.42$\pm$0.02 (0.44$\pm$0.03) & 0.29$\pm$0.04 & 0.25$\pm$0.06 & 0.35$^{+0.14}_{-0.13}$ & --- & --- & --- & 3.56 & 0.37$\pm$0.02 & 13.9/3 \nl 

\hline
 & & & & & & & & & & \nl 

NON-CF & & & & & & & & & & \nl 

A119           & 0.22$\pm$0.07 & 0.34$\pm$0.05 & 0.27$\pm$0.05 & 0.24$^{+0.07}_{-0.06}$ & 0.07$^{+0.06}_{-0.07}$ & --- & --- & 2.94 & 0.24$\pm$0.03 & 11.2/4 \nl 

A754           & 0.31$\pm$0.04 & 0.31$\pm$0.03 & 0.20$\pm$0.03 & 0.21$\pm$0.05 & 0.21$\pm$0.05 & --- & --- & 3.78 & 0.26$\pm$0.02 & 9.6/4 \nl 

A1656 (Coma)   & 0.18$\pm$0.04 & 0.28$\pm$0.03 & 0.24$\pm$0.03 & 0.22$\pm$0.03 & 0.25$\pm$0.03 & 0.19$\pm$0.04 &0.16$\pm$0.05 & 3.74 & 0.23$\pm$0.01 & 6.7/6 \nl 

A2256          & 0.34$\pm$0.04 & 0.29$\pm$0.03 & 0.22$\pm$0.03 & 0.21$\pm$0.05 & 0.25$\pm$0.05 & --- & --- & 3.26 & 0.26$\pm$0.02 & 7.5/4 \nl 

A2319          & 0.29$\pm$0.06 & 0.34$\pm$0.05 & 0.30$\pm$0.10 & 0.23$^{+0.14}_{-0.15}$ & 0.31$^{+0.11}_{-0.10}$ & 0.29$\pm$0.18 & --- & 3.86 & 0.31$\pm$0.03 & 0.8/5 \nl 

A3266          & 0.28$\pm$0.06 & 0.27$\pm$0.05 & 0.16$\pm$0.06 & 0.13$\pm$0.07 & 0.22$\pm$0.07 & --- & --- & 3.71 & 0.22$\pm$0.03 & 4.6/4 \nl 

A3376          & 0.26$^{+0.11}_{-0.10}$ & 0.30$^{+0.08}_{-0.07}$ & 0.17$\pm$0.07 & 0.32$^{+0.11}_{-0.10}$ & 0.13$^{+0.10}_{-0.09}$ & --- & --- & 2.46 & 0.23$\pm$0.04 & 3.4/4 \nl 

A3627          & 0.10$\pm$0.08 & 0.30$\pm$0.07 & 0.31$\pm$0.06 & 0.27$\pm$0.07 & 0.25$\pm$0.05 & 0.26$\pm$0.07 & 0.37$\pm$0.09 & 3.09 & 0.27$\pm$0.02 & 5.8/6 \nl 
\enddata

\tablenotetext{a}{Southern subcluster is excluded from 
the spectral analysis.}
\tablenotetext{b}{AGN at $\sim 4 ^\prime$ from the 
cluster center is excluded from the analysis.}
\tablenotetext{c}{A3562 hosts a modest cooling flow with 
a very small mass deposition rate of 
$37^{+26}_{-27}$ M$_\odot$ yr$^{-1}$ (Peres et al. 1998).}

\end{deluxetable}

\end{document}